\documentclass[a4paper,twocolumn,amsmath,amssymb,prl,longbibliography]{revtex4-1}
\usepackage{bm}
\usepackage{graphicx,color}
\usepackage{soul}
\usepackage{siunitx}
\usepackage{float}
\usepackage[colorlinks,linkcolor=magenta,citecolor=magenta]{hyperref}

\newcommand{\dd}{\mathrm{d}}

\newcommand{\pd}[2]{\frac{\partial #1}{\partial #2}}

\newcommand{\IInt}[3]{\int_{#2}^{#3}\dd #1\;}

\renewcommand{\vec}[1]{\mathbf #1}

\newcommand{\vhi}{\varphi}

\newcommand{\msig}{\bm\sigma}

\newcommand{\x}{\vec r}

\newcommand{\tx}{\tau_\text{r}}

\newcommand{\kT}{k_\text{B}T}
\newcommand{\eff}{\text{eff}}
\newcommand{\id}{\mathbf 1}

\AtBeginDocument{%
    \newwrite\bibnotes
    \def\bibnotesext{Notes.bib}
    \immediate\openout\bibnotes=\jobname\bibnotesext
    \immediate\write\bibnotes{@CONTROL{REVTEX41Control}}
    \immediate\write\bibnotes{@CONTROL{%
    apsrev41Control,author="08",editor="1",pages="1",title="0",year="1"}}
     \if@filesw
     \immediate\write\@auxout{\string\citation{apsrev41Control}}%
    \fi
}%

\begin{document}
\title{Force generation in confined active fluids: The role of microstructure}
\author{Shuvojit Paul$^1$}
\author{Ashreya Jayaram$^2$}%
\author{N Narinder$^1$}%
\author{Thomas Speck$^2$}%
\author{Clemens Bechinger$^1$}%
%\email{}
% \email{}
\affiliation{%
$^1$Fachbereich Physik, Universität Konstanz, 78464 Konstanz, Germany}%
\affiliation{%
$^2$Institut für Physik, Johannes Gutenberg-Universität Mainz, 55128 Mainz, Germany}%

\date{\today}

\begin{abstract}
We experimentally determine the force exerted by a bath of active particles onto a passive probe as a function of its distance to a wall and compare it to the measured averaged density distribution of active particles around the probe. Within the framework of an active stress, we demonstrate that both quantities are - up to a factor - directly related to each other. Our results are in excellent agreement with a minimal numerical model and confirm a general and system-independent relationship between the microstructure of active particles and transmitted forces.
  
\end{abstract}

\maketitle

Depletion forces often dominate the effective forces amongst (colloidal) particles and their interactions with walls when suspended in fluids containing additional depletion agents, e.g. a background of smaller particles, non-adsorbing polymers, micelles or vesicles~\cite{asakura1954interaction,mao@1995depletion,gotzelmann98}. At sufficiently small colloid-colloid or colloid-wall distances, excluded volume and other effects lead to anisotropic distributions of background particles around the colloids causing attractive and repulsive depletion forces.
Depletion forces are also generated by active particles (APs) that are capable of self-propulsion, e.g., motile bacteria, molecular motors, spermatozoa or active colloids (for a review see~\cite{bechinger2016active}). Despite noticeable similarities, such $\emph{active}$ depletion forces lead to an even richer behavior including long-ranged oscillatory forces onto confining walls (the pressure)~\cite{angelani2011effective,ray14a,ni15a,yan15,solon2015pressure,leite16a,junot2017active}, bath-mediated interactions between inclusions~\cite{yamchi17,liu2020constraint,feng21}, non-Gaussian diffusion of immersed colloidal particles~\cite{wu00,ortlieb19}, and the unidirectional motion of asymmetric objects in the presence of bacterial baths~\cite{angelani2009micromotor,di2010bacterial,sokolov2010swimming,mallory2014curv}. Opposed to passive, i.e., Brownian depletion agents, where a close relationship between their microstructure and the resulting depletion force has been established~\cite{asakura1954interaction,mao@1995depletion,gotzelmann98,Royall2005fluid,Kleshchanok2008direct,Lekkerkerker2011}, however, until now such a direct connection between the non-equilibrium steady state distribution of APs and the active depletion forces has not been reported in experiments.

In general, the mechanical force exerted on a probe particle is given by
\begin{equation}
  \vec F = \oint_{\partial\mathcal A}\dd\ell\;\vec n\cdot\msig
  \label{eq:f}
\end{equation}
where $\msig(\x)$ is the local stress in the bath due to the probe and walls
% generated by the depletion agent 
and $\partial\mathcal A$ is an arbitrary boundary with normal vector $\vec n$ enclosing the probe. In passive baths, the stress is directly related to the equilibrium distribution of depletants. For hard particles, highly accurate approximations have been constructed but this approach also extends to other interactions~\cite{attard89,gotzelmann98}.

Here we apply the theoretical framework of an ``active stress''~\cite{speck20a,speck21a,speck21b} to an experimental system to obtain the force exerted by APs onto a probe particle within a circular cavity from the measured AP distribution around the probe. The results are in excellent agreement with independent force measurements using optical tweezers and are also corroborated by numerical simulations. Notably, our approach does not require knowledge of the specific interactions between APs and the probe thereby facilitating the application of this concept to many other synthetic or living active systems. 

\begin{figure}[b!]
    \centering
    \includegraphics[scale=0.27]{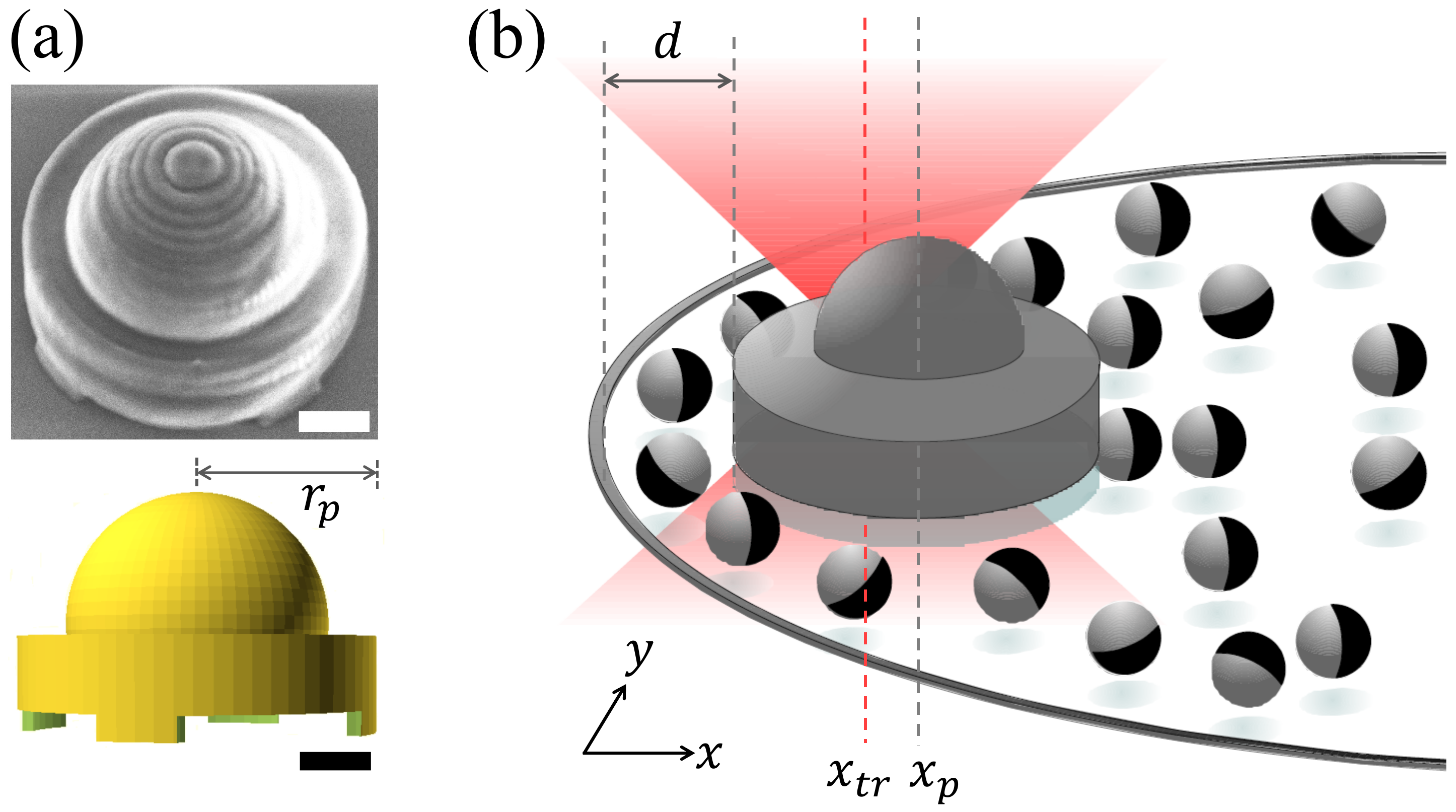}
    \caption{(a) Electron scanning microscope image (top) and schematic picture (bottom) of the probe particle used. The scale bars are $3~\mu$m each. (b) Schematic illustration of the experimental situation. The probe particle at $x_p$ and surface-to-surface distance to the circular wall $d$ is trapped by an optical tweezers ($x_{tr}$ denotes the center of the trap). The smaller spheres with the dark caps correspond to the APs.}
    \label{fig:schematic}
\end{figure}

Active colloidal particles are made by evaporating $20$~nm thin carbon caps onto silica spheres with diameter $2r_{a} \approx 2~\mu$m. They are suspended in a mixture of water and propylene glycol $n$-propyl ether (PnP) (0.4 mass fraction of PnP). Under laser illumination ($\lambda=532$~nm, $I \approx 7~\mu$W$/\mu$m$^{2}$), the fluid near the light-absorbing caps locally demixes leading to compositional surface flows and thus to self-propulsion of the particles~\cite{gomez2017tuning,ruben2020transient}. Owing to gravity and hydrodynamic interactions with the lower sample plate, the translational and rotational motion is effectively restricted to two dimensions (2D)~\cite{gomez2017tuning}. A suspension of APs with area fraction $0.31\pm0.02$ is laterally confined to a lithographically fabricated circular cavity with $100~\mu$m diameter. For the slow propulsion velocity $v=0.40\pm0.08~\mu$m/s used in this study, no motility-induced phase separation, i.e., formation of particle clusters, is observed~\cite{buttinoni2013dynamical,jeremie2013}. Under such conditions, the demixing zones of the binary fluid remain rather close to the AP surfaces which largely reduces phoretic interactions between the particles.
From the APs' rotational diffusion time $\tx\approx 23.7$~s, one obtains their persistence length $\ell=v\tx \approx 9.5~\mu$m (\emph{Supplemental Information}~\cite{sm}). For the measurement of active depletion forces, specifically designed probe particles were fabricated via a two-photon laser writing process~\cite{anscombe2010direct}. They consist of discs with diameter $2r_{p}=15~\mu$m and thickness $3.5~\mu$m with four legs underneath to minimize the friction with the substrate [Fig.~\ref{fig:schematic}(a)]. On top of the discs we designed hemispheres with diameter $11~\mu$m to guarantee a harmonic optical trapping potential when a vertically incident laser beam ($\lambda=1064$ nm) is focused onto the probe. The chosen intensity and focus diameter of the laser yield an optical trap stiffness $k=0.50\pm0.02~$pN/$\mu$m (\emph{Supplemental Information}~\cite{sm}). The force exerted on the probe is then obtained from the probe's mean displacement relative to the trap center $\Delta x = x_{p}-x_{tr}$ [Fig.~\ref{fig:schematic}(b)]. Prior to each force measurement, we allow the system to equilibrate for $\approx15$ minutes with the AP's activity turned off. After illuminating the sample, we wait $\approx10$ minutes to reach a steady state before recording the position of the probe and APs for $2000$~s using a center-of-mass tracking algorithm at a frame rate of $2$ Hz.

\begin{figure}[t!]
    \centering
    \includegraphics[scale=0.35]{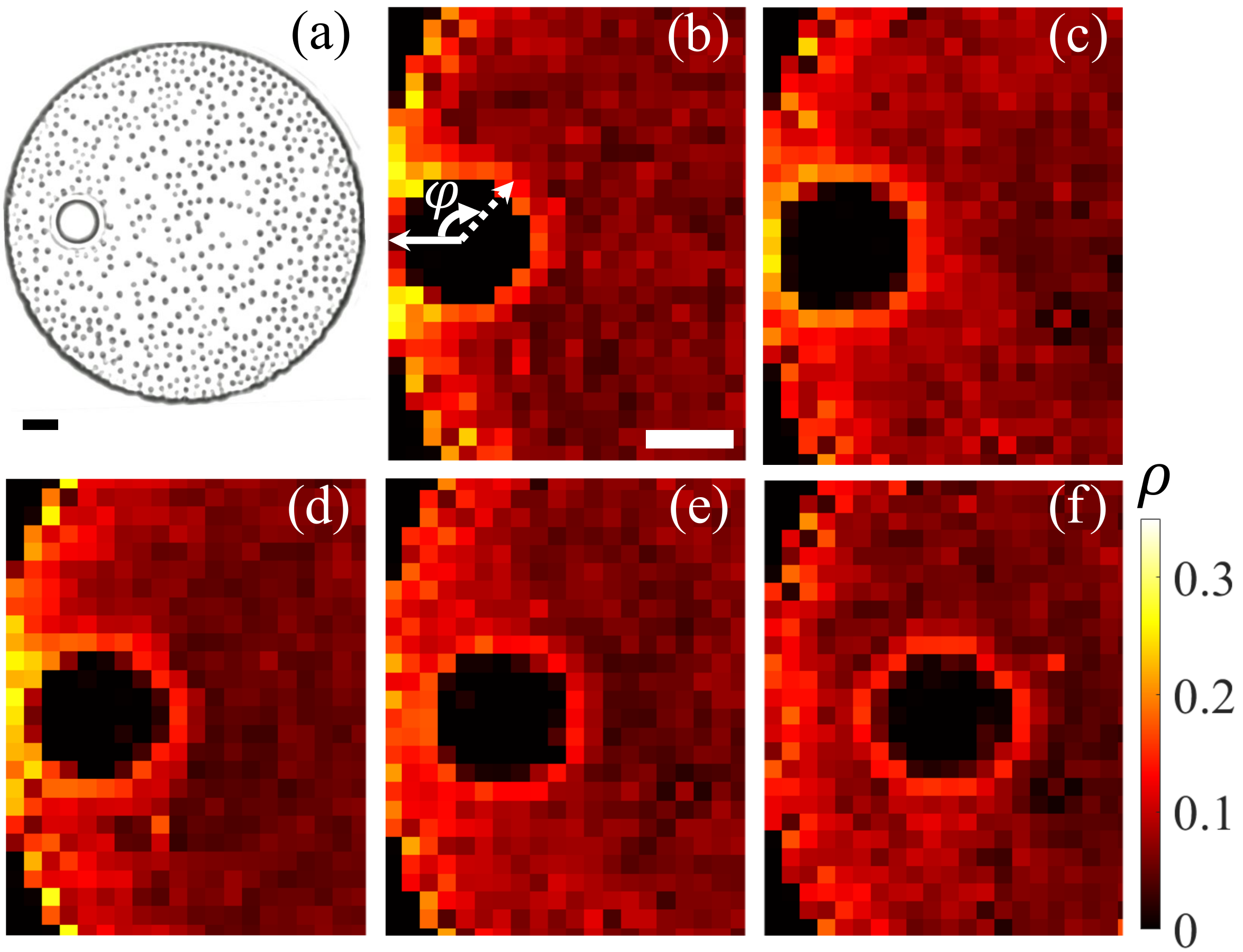} 
    \caption{(a) Experimental snapshot showing APs (dark points) and the optically trapped probe particle inside the circular cavity. Measured AP density distributions for different probe-wall distances (b) $d'=0.46$, (c) $1.02$, (d) $1.59$, (e) $2.78$, (f) $6.30$. Scale bars are $10~\mu$m each.}
    \label{fig:experiment}
\end{figure}

\begin{figure}[t!]
    \centering
    \includegraphics[scale=0.24]{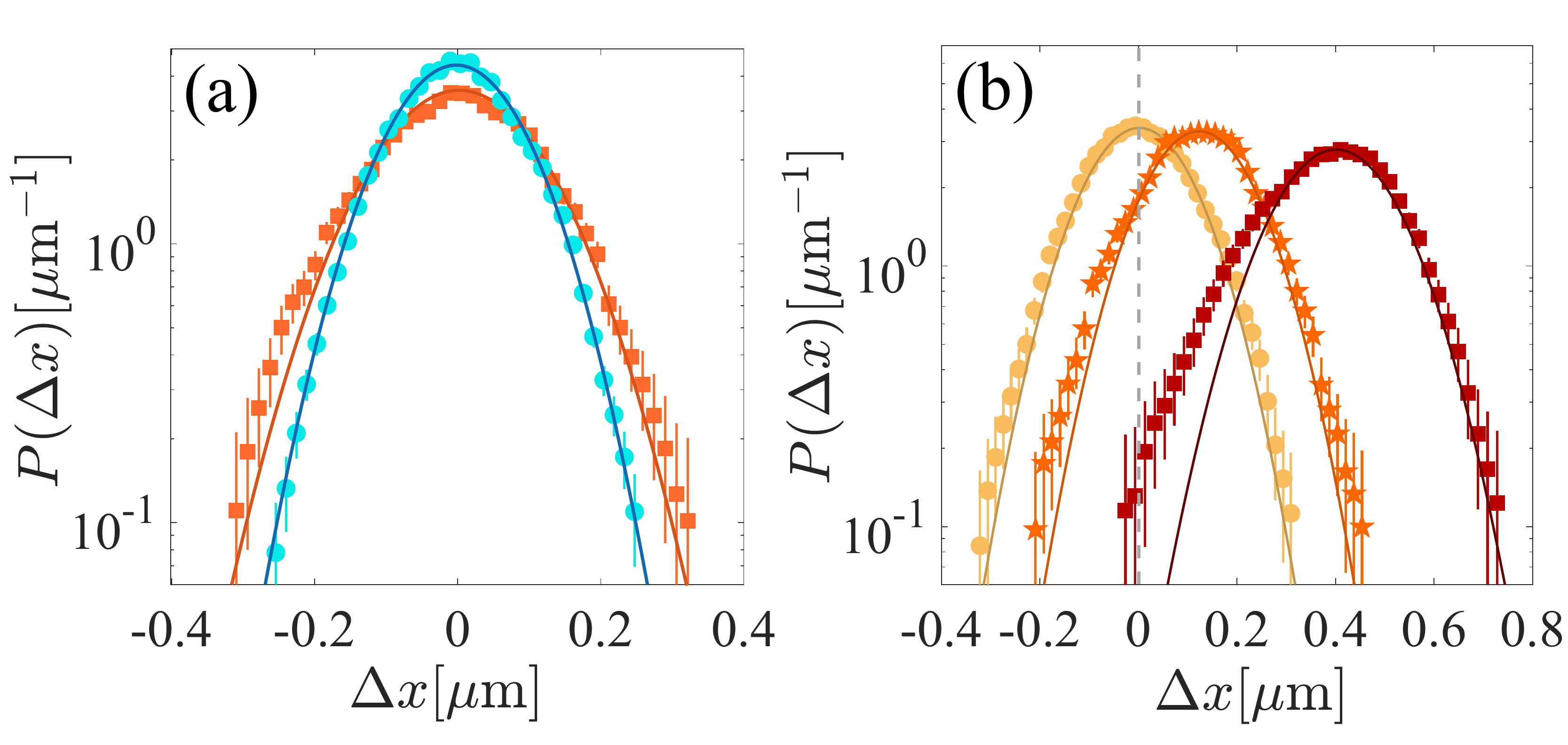} 
    \caption{(a)~Distribution $P(\Delta x)$ at the center of the circular confinement in the presence of active (squares) and passive (circle) particles. The solid lines are Gaussian fits. (b)~Distribution $P(\Delta x)$ with the probe at $d'=1.02$ (stars), $1.59$ (squares) and $6.30$ (circles). The solid lines correspond to fits with a Gaussian distribution. The error bars show the standard deviation of experimental data calculated over five independent measurements.}
    \label{fig:probability}
\end{figure}

Figure~\ref{fig:experiment}(a) shows a typical snapshot of the experiment. From the time-averaged configurations we calculate the spatial density $\rho(x,y)$ of APs, which is plotted in Figs.~\ref{fig:experiment}(b--f) for different probe-wall distances normalized by the AP diameter, i.e., $d'=d/(2r_{a})$. In agreement with previous studies, as a result of the APs' finite reorientation dynamics, their density is increased at the surfaces of the cavity wall and the probe~\cite{schaar2015detention}. In particular at small $d'$, APs strongly accumulate in the wedges formed by the probe and the wall~\cite{leite16a}. This asymmetry in the AP distribution leads to $d'$-dependent changes of the probe's fluctuations relative to the optical trap center as shown by the corresponding probability distributions $P(\Delta x)$. Figure~\ref{fig:probability}(a) corresponds to the situation where the probe is near the center of the cavity. 
Regardless of whether the colloidal particles are passive or active (i.e. green laser illumination is off or on), $P(\Delta x)$ is symmetric around $\Delta x =0$ indicating the absence of an effective force (solid lines). However, compared to passive particles where $P(\Delta x)$ agrees well with a Gaussian distribution, in case of an active background $P(\Delta x)$ becomes broader and exhibits non-Gaussian tails. Such tails are caused by the finite persistence length of APs and are in good agreement with other studies using active baths~\cite{maggi14a,argun2016pre}. When the probe is positioned closer to the cavity wall, however, the maximum of the distributions shift to the right relative to $\Delta x=0$ and the distributions become asymmetric [Fig.~\ref{fig:probability}(b)], both indicating an additional force acting on the probe.

\begin{figure}[t!]
    \centering
    \includegraphics[scale=0.40]{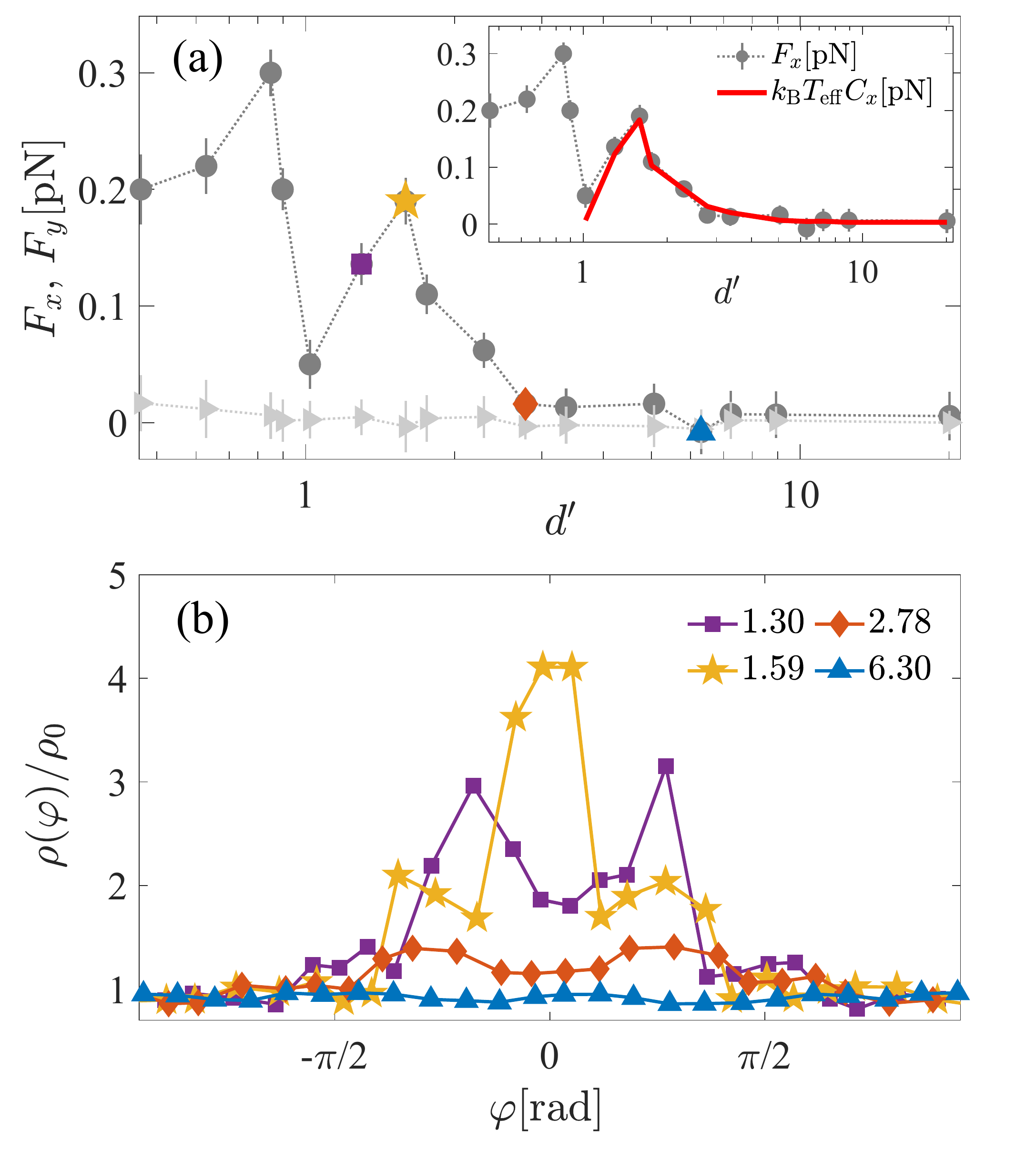}
    \caption{(a)~Experimentally measured values of  $F_x$ (dark symbols) and $F_y$ (light symbols) as a function of normalized distance $d'=d/(2r_a)$ of the probe from the wall of the circular confinement with radius $r_\text{con}=50~\mu$m. The error bars correspond to the standard deviations calculated over five independent measurements. The inset shows $F_x$ (symbols) together with the prediction according to Eq.~\eqref{eq:f:eff} from the measured number density $\rho$ (red solid line). (b)~Density $\rho(\varphi)$ of APs (normalized by the global number density $\rho_{0}$) around the probe for different values $d'$ (being labeled with identical symbols (colors) as in (a).}
   \label{fig:exp_panel}
\end{figure}

From the probe's mean displacement relative to the optical trap's center in $x$ and $y$ direction, we obtain the force $\vec F=(F_x,F_y)$ acting on the probe, which is shown as symbols in Fig.~\ref{fig:exp_panel}(a) as a function of $d'$. In particular at short distances, $F_x(d')$ displays a strong non-monotonic behavior and eventually decays to zero at large distances $d'$. Such behavior is in qualitative agreement with repulsive active forces observed in previous experimental and theoretical studies~\cite{angelani2011effective,ni15a,yamchi17,duzgun2018active,liu2020constraint}. $F_x$ becomes zero at distances $d'\gtrapprox 5$ corresponding approximately to the APs' persistence length. Due to symmetry reasons, $F_{y}$ is essentially zero and only fluctuates slightly around zero.

A qualitative understanding of the non-monotonic and repulsive behaviour of $F_{x}(d')$ is immediately obtained from the density distributions shown in Figs.~\ref{fig:experiment}(b--f). At a probe-wall distance $d'=0.46$, APs are unable to pass the probe-wall spacing, which leads to a strong increase of $\rho$ in the wedges formed by the probe and the wall. This density inhomogeneity around the probe causes a large repulsive active depletion force $F_{x}$ acting on the probe [Fig.~\ref{fig:experiment}(b)] in stark contrast to attractive forces generated by passive depletion agents at such small distances. Upon slightly increasing $d'$, APs are able to ``squeeze'' through the space between the probe and the wall, thereby further pushing the probe away from the wall. When $d'$ becomes comparable to the AP diameter, this squeezing-induced effect disappears and the AP density enrichment near the wall is much less pronounced [Fig.~\ref{fig:experiment}(c)]. This explains the appearance of the first maximum of $F_{x}$ [Fig.~\ref{fig:exp_panel}(a)]. Because APs generally tend to accumulate near walls, for $d'>1$, on average, a dense monolayer forms at the circular cavity wall and the above discussed accumulation of APs at the wedges will repeat in a similar fashion at larger distances. This is shown for $d'=1.59$ where the density of APs between the probe and the wall is again increased [Fig.~\ref{fig:experiment}(d)] which rationalizes the second peak in $F_{x}$ [Fig.~\ref{fig:exp_panel}(a)]. For larger distances, the density around the probe becomes more and more homogeneous, which eventually leads to the disappearance of an active depletion force [Figs.~\ref{fig:experiment}(e,f)].

As a first step towards a quantitative relationship between the AP distribution and the active depletion force, we determine from $\rho(x,y)$ the angle-resolved AP density distribution $\rho(\varphi)$ around the probe particle with $\varphi$ defined in Fig.~\ref{fig:experiment}(b) (for details regarding the determination of $\rho(\varphi)$ see \emph{Supplemental Information}~\cite{sm}). Figure~\ref{fig:exp_panel}(b) shows $\rho(\varphi)$ normalized by the total particle density $\rho_0$ for several values of $d'$ [marked as colored symbols in Fig.~\ref{fig:exp_panel}(a)]. Clearly, the non-monotonic behavior in $F_{x}(d')$ is also reflected in the corresponding $\rho(\varphi;d')$, suggesting a close link between the two quantities.

To yield a direct relationship between the active depletion force and the AP distribution, we write the force on the probe as $\vec F=\int_{\mathcal A}d^2\x\;\rho\nabla u$ with local density $\rho(\x)$ of active bath particles, and $u(r)$ the pair potential characterizing the bath-probe interaction as a function of their (center-to-center) distance $r$. The force thus depends on the averaged AP configuration given by their density profile with respect to the probe center [cf.~Fig.~\ref{fig:exp_panel}(b)]. In contrast to a dilute \emph{passive} bath in thermal equilibrium, where $\rho(r)\propto\text{e}^{-u(r)/(\kT)}$, such a relationship is no longer valid in case of an active bath~\cite{ginot15} and thus fails to predict the force $\vec F$~\cite{liu2020constraint}.

Under steady state conditions and regardless of whether the system is in thermal equilibrium or not, force balance dictates the validity of the hydrostatic equation $\nabla\cdot\msig-\rho\nabla u=0$ (in the absence of particle currents~\cite{speck21a}). Exploiting the divergence theorem leads to Eq.~\eqref{eq:f}. Importantly, if we choose a contour $\partial A$ outside the range of $u(r)$, the interactions $u(r)$ do not show up explicitly in the expression for the local stress~\cite{speck20a} 
\begin{equation}
  \msig = -\rho\kT\id - \kT v\tx\left[\frac{\hat v}{2D_0}\rho\id-(\nabla\vec p)^{ST}\right]
  \label{eq:sig}
\end{equation}
even though they do shape the distribution of APs around the probe. Here, the local polarization field $\vec p(\x)$ of APs enters.
Moreover, the local propulsion speed $\hat v(\x)$ is reduced compared to the bare propulsion speed $v$ due to interactions with neighboring particles. The symmetric and traceless derivative is $(\nabla\vec p)^{ST}=\partial_ip_j+\partial_jp_i-(\nabla\cdot\vec p)\delta_{ij}$.
The view expressed in Eq.~\eqref{eq:sig} relates the non-vanishing polarization due to the aggregation at the probe and confinement to an active stress tensor, but alternative views have been proposed in which the polarization is identified with an external one-body force~\cite{omar20}. However, this distinction neither changes the force balance nor the mechanical force $\vec F$ on the probe.

We now consider a circular contour $\partial\mathcal A$ with radius $r_b>r_p+r_a$ and normal vector $\vec n=\vec e_r$ given by the radial unit vector. We choose $r_b$ as close to the probe as possible but outside the range of the pair potential, $u(r>r_b)\approx 0$. Plugging Eq.~\eqref{eq:sig} into Eq.~\eqref{eq:f}, we obtain (for details, see \emph{Supplemental Information}~\cite{sm})
\begin{multline}
  \vec F = -\kT\left(1+\frac{v\tx v_\text{eff}}{2D_0}\right) \IInt{\vhi}{0}{2\pi}r_b\vec e_r\rho(r_b,\vhi) \\ + \kT v\tx r_b\left[\pd{}{r}\IInt{\vhi}{0}{2\pi}\vec p(r,\vhi)\right]_{r=r_b}.
\end{multline}
The second integral involving the polarization can be eliminated by setting $\hat v\vec p=D_0\nabla\rho$ exploiting the fact that our symmetric probe does not generate a particle current. Moreover, symmetry dictates that $F_y=0$ [cf.~Fig.~\ref{fig:exp_panel}(a)]. If we further use that the density decays exponentially away from the probe we obtain the rather simple result~\cite{sm}
\begin{equation}
  F_x = \kT_\text{eff}C_x
  \label{eq:f:eff}
\end{equation}
for the force, whereby the integral $C_x(r_b)=-\IInt{\vhi}{0}{2\pi}r_b\cos\vhi\rho(r_b,\vhi)$ only depends on the density of APs outside the probe. While Eq.~\eqref{eq:f:eff} resembles the passive result with an elevated effective temperature $T_\text{eff}(r_b)$, the distribution of APs around the probe markedly deviates from the equilibrium profile. The temperature $T_\text{eff}$ simply encodes the elevated density fluctuations due to the activity of the particles.

The solid line in the inset of Fig.~\ref{fig:exp_panel}(a) shows the predicted force $F_x$ calculated from the experimentally measured density profiles with $r_b=r_p+2.2r_a$. We observe excellent agreement with the independently measured forces with an effective temperature $T_\text{eff}\simeq 17T$ for all distances $d'$. Note that for distances $d'<1$ the integration contour would intersect the wall which does not allow the evaluation of forces in this range. We remark that $\kT_\text{eff}$ and $C_x$ individually depend on $r_b$ but $F_x$ is independent of the precise choice of the integration contour.

\begin{figure}[t!]
    \centering
    \includegraphics[scale=0.4]{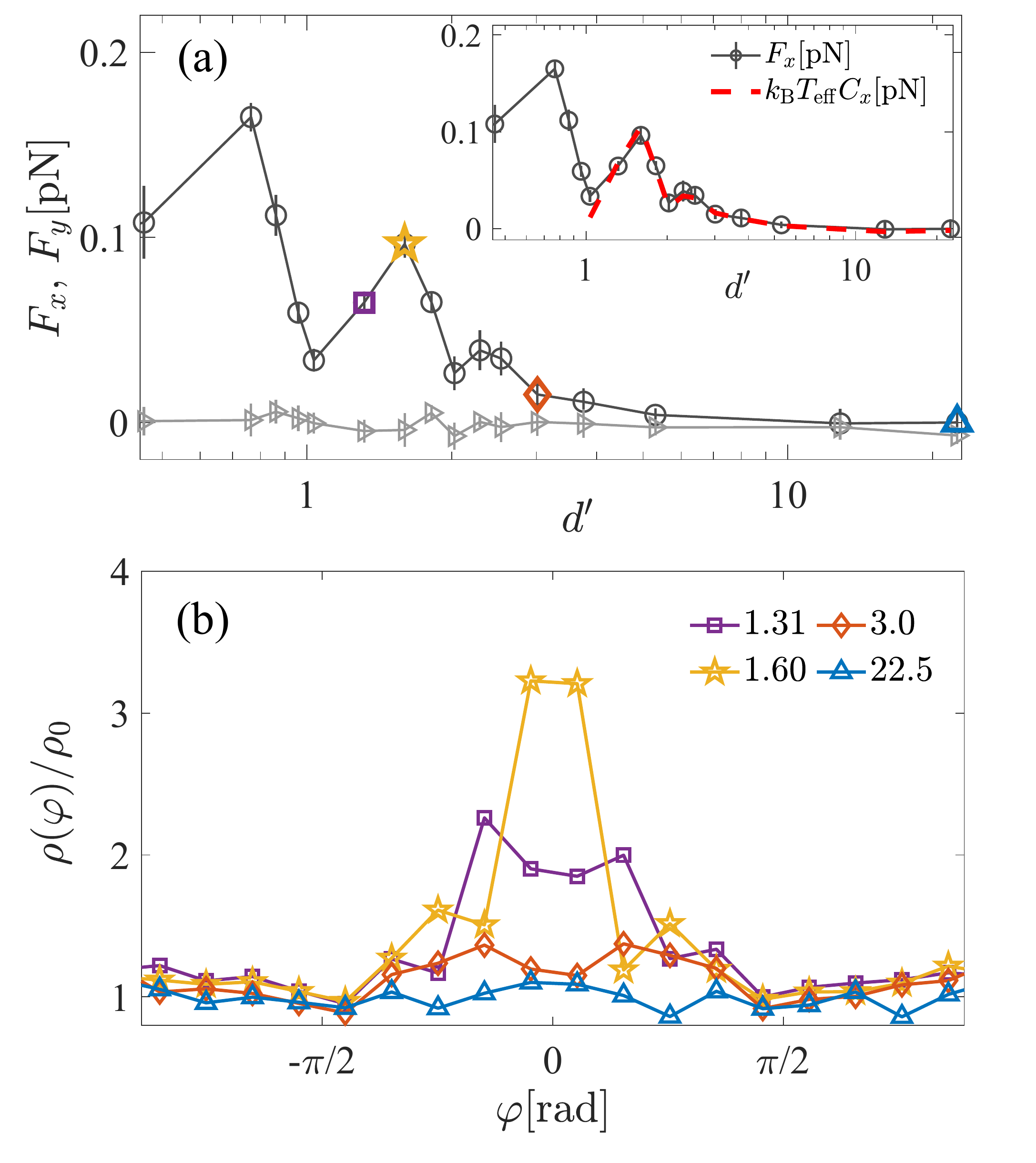}
    \caption{Simulation results for active Brownian particles. (a)~Force on the probe in $x$ (open circles) and $y$ (open triangles) direction as a function of normalized distance $d'$ for a circular confinement of radius $r_\text{con}=50~\mu$m. The error bars correspond to the standard deviation calculated over five independent simulation runs. The inset shows the measured force $F_x$ (symbols) together with the prediction Eq.~\eqref{eq:f:eff} from the measured number density $\rho$ (red line). (b)~Density $\rho(\vhi)$ of APs (normalized by the global number density $\rho_{0}$) around the probe on a circle of radius $r_b=r_p+2.2r_a$ for different $d'$. $F_{x}$ corresponding to these $\rho(\vhi)$ shown in (b) is marked with the same symbols in (a).}
   \label{fig:sim-force}
\end{figure}

We corroborate our results with numerical simulations employing a minimal model of active Brownian particles. Within this model, hydrodynamic interactions are neglected and only the excluded volume of APs and their self-propulsion with constant speed $v$ is considered. All model parameters are fixed by the experimental values. Further details are provided in the \emph{Supplemental Information}~\cite{sm}. Figure \ref{fig:sim-force}(a) shows the force $\vec F$ as a function of $d'$ obtained from simulations. As in the experiments [cf.~Fig.~\ref{fig:exp_panel}(a)], the force experienced by the probe is always repulsive and oscillates before eventually decaying to zero. In addition, the positions of the maxima agree very well with experiments. The inset of Fig.~\ref{fig:sim-force}(a) shows the measured force together with the prediction of Eq.~\eqref{eq:f:eff}, which also for this model system shows excellent agreement with a single elevated temperature $T_\eff\simeq 16.25T$ independent of the probe-wall separation.

The only difference lies in the magnitude of $\vec F$ and the angular densities $\rho(\varphi)$ [Fig.~\ref{fig:sim-force}(b)] which are consistently smaller in simulations. To understand those differences in more detail, we have determined the distribution of contact times, i.e. the times APs remain near the lateral confinement and the probe particle, respectively. We find that using an entirely repulsive model underestimates the width of the experimentally determined distribution of contact times, see \emph{Supplemental Information}~\cite{sm}. Adding an attraction to the model that effectively accounts for, e.g., phoretic attractions and the slowing down of AP velocities due to surface roughness or hydrodynamic effects yields better consistency with the experimental data and concurrently increases the force magnitude, reaching good quantitative agreement~\cite{sm}.

In summary, we have demonstrated with experiments and simulations that the forces exerted by an active bath onto a probe particle can be obtained from the AP density around the probe employing the recently proposed concept of active stress. Remarkably, this approach does not require explicit knowledge of system-specific AP-probe interactions; therefore, it is applicable to many other active systems. In addition to understanding the motion of Brownian objects in baths of synthetic active particles, our work may also elucidate the role of active depletion forces in living systems which are believed to contribute to the organization of intra-cellular structures~\cite{almonacid2015active,Razin2017archimeded}.

\begin{acknowledgments}
We thank J.~Steindl for the fabrication of probe particles. S.P. acknowledges support from the Alexander von Humboldt foundation, A.J. and T.S. for funding from the Deutsche Forschungsgemeinschaft through TRR 146 and C.B. for financial support from the ERC Advanced Grant ASCIR (Grant No.~693683).
\end{acknowledgments}

% -- bibliography --
%\bibliography{references}
%%%%%% bibliography in the main file %%%%%%


\end{document}